\documentstyle[preprint,aps,epsfig]{revtex}
\newif\iftightenlines\tightenlinesfalse
\tightenlines\tightenlinestrue

 \def\to{\rightarrow}

\def \Oi{{\mathcal O}}
\def \tm{{\tilde m}}
\def \be{\begin{equation}}
\def \ee{\end{equation}}
\def \bea{\begin{eqnarray}}
\def \eea{\end{eqnarray}}

\begin{document}
\preprint{\vbox{\baselineskip=14pt%
}}

\title{Chargino Contributions in $B \rightarrow \phi K_S$ CP Asymmetry}
\author{Yili Wang}
\address{
Department of Physics and Astronomy,
Iowa State University,
Ames, Iowa 50011, USA.
}
\date{\today}
\maketitle
\begin{abstract}
CP asymmetry in $B \to \phi K_S$ decay  is studied in 
the special context of supersymmetry theories, in which the charginos play 
an important role. We find that in addition to the gluino, the chargino can 
also make large contributions to CP asymmetry in $B \to \phi K_S$ decay. 
After considering the constraints from $B \to J/\psi K_S$ decay, we study three
special scenarios:  (a). Large mixing between left-handed charm and 
top squarks(LL mixing); 
(b). Large mixing between right-handed charm and top squarks (RR mixing); 
(c). Large mixing between both left-handed charm-top squark and  right-handed 
charm-top squark(LL+RR mixing). We show quantitatively that because of large 
squark mixing within the second and third generations, an $\Oi$(1) effect 
on CP violation in $B \to \phi K_S$ is possible. 

{\noindent\scriptsize{ PACS numbers: 13.25.Hw; 12.60.Jv; 11.30.Pb}}
\end{abstract}

\pagebreak

\section{Introduction} \label{intro}

In standard model, all the  CP asymmetries in the B system can be 
accounted for with a single phase in Cabbibo-Kobayashi-Maskawa 
mixing matrix\cite{ckm}. It predicts that the time dependent CP asymmetry in 
$B \to \phi K_S$ decay is the same as the  CP asymmetry in $B \to J/\psi K_S$ 
decay. That is: 
$$S_{J/\psi K_S} = S_{\phi K_S} = sin 2\beta = 0.731 \pm 0.056 (SM)$$  
Any difference between these two decays would be evidence of new physics 
beyond the standard model. Finding CP violation in the B system is a very 
important goal in B factories.

BaBar\cite{babar} and Belle\cite{belle}  measurements on CP asymmetries in 
$B \to J/\psi K_S$ decay have provided the first evidence for CP violation 
in the B system.
\begin{eqnarray}
\begin{array}{cccc}
sin 2\beta(B \to J/\psi K_S) & = & 0.741 \pm 0.067 \pm 0.033 & (BaBar) \cr
sin 2\beta(B \to J/\psi K_S) & = & 0.733 \pm 0.057 \pm 0.028 &  (Belle) \cr  
\end{array}
\end{eqnarray}
 The average of their results is: $$S_{J/\psi K_S} = sin 2\beta = 0.734 \pm 0.054,$$ This result is in good agreement with the SM prediction \cite{jp}.  Therefore, we can conclude that  CP is significantly violated and the CKM phase angle is the dominant source of CP violation.

However, the CKM mechanism is challenged by the recent measurements of 
the BaBar and Belle collaborations in $B \to \phi K_S$ decay.
\begin{eqnarray}
\begin{array}{cccc}
sin 2\beta(B \to \phi K_S) & = & +0.45 \pm 0.43 \pm 0.07 & (BaBar) \cr
sin 2\beta(B \to \phi K_S) & = & -0.96 \pm 0.50_{-0.11}^{+0.09} &  (Belle)\cr   
\end{array}
\end{eqnarray}
 The average of these two measurements implies: 
$$S_{\phi K_S} = sin2\beta = -0.15 \pm 0.33.$$ which deviates from the 
SM prediction by 2.7$\sigma$. The $2\sigma$ difference between $S_{J/\psi k_S}$ and $S_{\phi K_S}$ is expected to be an indication of new physics. Many papers 
have explored potential new physics 
contributions \cite{new3,new1,new2,new4,new5} to $B \to \phi K_S$ decay. 
Among them, supersymmetry is one of the most significant candidates to 
account for this $2\sigma$ deviation.

In $B \to J/\psi K_S$ decay, standard model contributions dominate the tree 
level diagrams and any new physics enters only in loop diagrams, 
which are suppressed naturally. As a result, in order to have a significant supersymmetric contribution, a large SUSY CP violating phase is required. Such large phases usually do not exist in most of the supersymmetric models. 
Therefore the CP asymmetry in $S_{J/\psi K_S}$ is dominated by the standard 
model contribution. Unlike the case of $B \to J/\psi K_S$, 
there are no tree level 
diagrams in $B \to \phi K_S$ decay because $b \to s\bar s s$ , the dominant 
process in $B \to \phi K_S$, is induced only at one loop level in the 
standard model. Both SM and SUSY contributions appear at one loop level. 
Thus SUSY contributions are more significant in $B \to \phi K_S$ decay 
and are expected to explain the 2$\sigma$ difference.

SUSY contributions to $B \to \phi K_S$ decay come mainly from penguin and 
box diagrams containing gluino, chargino and neutralino loops.
The contributions from neutralino loops are much smaller than the 
contributions from gluino and chargino loops. 
We ignore contributions from neutralinos in our computation.  
There are several papers investigating on SUSY contributions to  
$B \to \phi K_S$ decay \cite{s1,s8,s2,s7,s6,s3,s9,s4,s5}, especially the gluino
loop contributions\cite{susy1,susy2,susy3,phi1,hm,phi2}. Only a few include 
the chargino diagrams, which can also affect the result substantially\cite{chi}.
The main purpose of this paper is to study chargino contributions to the
CP asymmetry in  $B \to \phi K_S$ decay. The chargino contributions depend 
on the details of the model. 
However, we try to be as general as possible in our analysis.
To emphasize the chargino loop contributions, we work in a special 
basis where the down-type squark mass matrix is diagonalized. 
There is no intergeneration mixing in the down-type squark sector. 
In this scenario, gluino contributions to $B \to \phi K_S$ decay are ruled out 
and chargino contributions are the dominant contributions. 

This paper is organized as follows: In the next section, we present the formula
for CP violation in the B system and introduce the effective field theory 
for the $b \to s $ transition. Section ~\ref{num} is devoted to the 
discussion of supersymmetric contributions to the CP asymmetry in 
$B \to \phi K_S$ process.  
Section ~\ref{con} includes the conclusion and the loop functions 
are showed in Appendix.
 
\section{CP Violation in $B \to \phi K_S$ Decay }\label{cp1}

In the B system, the time dependent CP asymmetry in $B \to \phi K_S$ 
decay is defined as
\bea
A_{\phi K_S}(t)&=&\frac{\Gamma (\overline{B}^0(t)\to\phi K_S)-\Gamma 
(B^0(t)\to\phi K_S)} {\Gamma (\overline{B}^0(t)\to\phi K_S)+\Gamma (B^0(t)
\to\phi K_S)} \nonumber 
\eea
\bea
&=&C_{\phi K_S}\cos\Delta M_{B_d}t+S_{\phi K_S}\sin\Delta M_{B_d}t
\eea

where $\Delta M_{B_d}$ is the mass difference between the two neutral B mesons.
$C_{\phi K_S}$ and $S_{\phi K_S}$ are
\begin{equation}
C_{\phi K_S}=\frac{|\lambda(\phi K_S)|^2-1}{|\lambda(\phi K_S)|^2+1}, 
\ \ \
S_{\phi K_S}=\frac{2Im \left[\frac{q}{p}~\lambda(\phi K_S)\right]} 
{|\lambda(\phi K_S)|^2+1}.  \label{eq:SCdef}
\end{equation}

$p$ and $q$ are the mixing parameters defined as: $$|B_\pm> = p|B^0> \pm q|\bar{B}^0>$$

$\lambda_{\phi K_S}$ is a ratio of decay amplitudes:
\begin{equation}
\lambda_{\phi K_S}=\frac{\overline{A}(\overline{B}^0 \to\phi K_S)}{A({B}^0 \to\phi K_S)} \nonumber
\end{equation}
 
From above formulae, we notice that there are two sources of CP violation 
\cite{so1}. One is from the mixing $|\frac{q}{p}| \not= 1$, the other is from 
the ratio of decay amplitudes $|\frac{\overline{A}}{A}| \not= 1$~\cite{jp,am}. 
In the SM, loop diagrams with W boson and top quark dominate the 
$b \to sss$ decay channel. In this channel, 
$\lambda_{\phi K_S} = \frac{\overline{A}^{SM}(\overline{B}^0\to\phi K_S)}{A^{SM}(B^0\to\phi K_S)} = \frac{V_{tb} V_{ts}^{*}}{V_{tb}^{*} V_{ts}} = 1$. 
There is no CP violation from the ratio of decay amplitudes.  Thus, CP asymmetry in 
$B \to \phi K_S$ decay comes mainly from mixing , 
which is the same with  $B \to J/\psi K_S$ decay. 

Supersymmetry can contribute to mixing, $\Delta B = 2$ transitions, and to decay 
$\Delta B = 1$ transitions. SUSY contributions to the decay amplitudes are 
through penguin and box diagrams \cite{pen}, while contributions to the mixing 
parameters are only through box diagrams. To concentrate on the chargino 
contributions, we choose a basis, in which the down-type squark mass is 
diagonalized. There is no intergeneration mixing for down-type sector. 
The gluino contributions are ruled out by the choice of basis. 
SUSY contributions are mainly chargino contributions.

The effective Hamiltonian in $\Delta B = 1$ transitions is:

\begin{equation} \label{effect}
H =  \sum_{i=1}^{6}\left [ C_i(\mu){\Oi_i}(\mu)\right ] + C_g \Oi_g
\end{equation}

As in Ref.\cite{hm,bobeth}, the operators we choose are:
\begin{eqnarray}\label{operatorbasis}
{\Oi_1}&=& (\bar{s}_{ \alpha} \gamma_\mu P_L c_{ \beta})
(\bar{c}_{ \beta} \gamma^\mu P_Lb_{ \alpha}),\nonumber\\
{\Oi}_2 &=& (\bar{s}_{ \alpha} \gamma_\mu P_L c_{ \alpha})
(\bar{c}_{ \beta} \gamma^\mu P_L b_{ \beta}),\nonumber\\
{\Oi}_3 &=& (\bar{s}_{ \alpha} \gamma_\mu P_L b_{ \alpha})
(\bar{s}_{ \beta} \gamma^\mu P_L s_{ \beta}),\nonumber\\
{\Oi}_4&=& (\bar{s}_{ \alpha} \gamma_\mu P_L b_{ \beta})
(\bar{s}_{ \beta} \gamma^\mu P_L s_{ \alpha}),\nonumber\\
{\Oi}_5&=& (\bar{s}_{ \alpha} \gamma_\mu P_L b_{ \alpha})
(\bar{s}_{ \beta} \gamma^\mu P_R s_{ \beta}),\nonumber\\
{\Oi}_6&=& (\bar{s}_{ \alpha} \gamma_\mu P_L b_{ \beta})
(\bar{s}_{ \beta} \gamma^\mu P_R s_{ \alpha}),\nonumber\\ 
{\Oi}_g & =& \frac{g_s}{8 \pi^2}m_b (\bar{s}_{\alpha} 
T_{\alpha \beta}^a \sigma_{\mu \nu}P_R b_{\beta}) G^{a \mu \nu}
\end{eqnarray}
where $\alpha$ and $\beta$ are color indices, $P_L = (1 - \gamma_5)/2$, and $\sigma_{\mu \nu} = \frac{i}{2}[\gamma^\mu, \gamma^\nu]$. 
The contribution from the operator 
${\Oi}_\gamma  = \frac{e}{8 \pi^2}m_b (\bar{s}_{\alpha} \sigma_{\mu \nu}P_R b_{\alpha})F^{\mu \nu}$  is very small and is ignored in this paper.  
Operators proportional to s-quark mass $m_s$ are also ignored because of the 
small s-quark mass.
 
In the SM, the Wilson coefficients $C_{3 \sim 6}$ in $B \to \phi K_S$ decay 
have been computed to NLL order, and $C_g$ has been computed to LL order. 
The values we used are in ref.~\cite{sm}. 

In SUSY, up-squark masses are taken to be non-degenerate. We use the mass 
eigenbasis in the computation. There are large mixing between the second 
and third generations in up-type sector.  The mass matrix takes the form:

\begin{equation}
\label{massmatrix}
\tilde{m}^2_{\tilde{u}}= \left(
\begin{array}{cccccc}
\tm_{Lu}^2 & 0 & 0 & 0 & 0 & 0  \\
  0  & \tm_{Lc}^2 & \tm_{Lct}^2 & 0 & 0 & 0\\
  0  & \tm_{Lct}^{*2} & \tm_{Lt}^2 & 0 & 0 & m_t A_t cot\beta\\
 0  & 0 & 0 & \tm_{Ru}^2 & 0 & 0\\
 0  & 0 & 0 & 0 &  \tm_{Rc}^2  & \tm_{Rct}^2\\
 0 & 0& m_t A_t cot\beta  & 0 & \tm_{Rct}^{*2} & \tm_{Rt}^2 
\end{array} \right).
\end{equation}

This mass matrix can be diagonalized by a unitary matrix $R_{\tilde u}$. That is:$$R_{\tilde u}^+ \tm_{\tilde u}^2 R_{\tilde u} = diag(\tm_{L1}^2, \tm_{L2}^2, \tm_{L3}^2, \tm _{R1}^2, \tm_{R2}^2, \tm_{R3}^2),$$ It is useful to define the $6 \times 3$ matrices such that $$ (\Gamma^{q_L})_{ai} = (R_{\tilde u})_{ai},\ \ \ \ (\Gamma^{q_R})_{ai} = (R_{\tilde u})_{a i+3}.$$ 
where a = 1, ... ,6 denotes the mass eigenstates 
and i = u,c,t labels the gauge eigenstates. 

The chargino mass matrix is
\begin{equation}
\label{charginomatrix}
\tilde{M}_{\tilde{\chi^\pm}}= \left(
\begin{array}{cc}
M_2 & \sqrt{2} M_W sin\beta  \\
 \sqrt{2} M_W cos\beta  & \mu
\end{array} \right).
\end{equation}
where we assume the mass of the W-ino field, $M_2$, is real and positive. 
The non observation of the neutron electric dipole moment (EDM) strongly 
constrains the phase of $\mu$. A non-zero phase of $\mu$ does not significantly affect our results, 
we keep the phase of $\mu$ to be zero for sake of simplicity.

The chargino mass matrix can be diagonalized by a biunitary transformation
$$ diag(m_{{\tilde \chi_1}^\pm}, m_{{\tilde \chi_2}^\pm}) = U \tilde{M}_{{\tilde \chi}^\pm} V^T $$ where $U$ and $V$ are unitary matrices defined as
\begin{equation}
\label{umatrix}
 U = \left(
\begin{array}{cc}
 cos \theta_U & sin\theta_U  \\
 -sin\theta_U  & cos\theta_U
\end{array} \right).
\end{equation}

\begin{equation}
\label{vmatrix}
V = \left(
\begin{array}{cc}
 cos\theta_V & sin\theta_V \\
 -sin\theta_V  & cos\theta_V
\end{array} \right).
\end{equation}

with mixing angles \cite{bobeth}

\bea
\label{eqn:mixing}
\sin 2\theta_{U,V} &=& \frac{2M_W\big[M_2^2+ \mu^2 \pm (M_2^2-\mu^2)\cos 2\beta + 2\mu M_2\sin2\beta]^{1/2}}{m^2_{\tilde{\chi}^{\pm}_1}-m^2_{\tilde{\chi}^{\pm}_2}}, \nonumber \\
\cos2\theta_{U,V} &=& \frac{M_2^2- \mu^2\mp 2M_W^2 \cos2\beta}{m^2_{\tilde{\chi}^{\pm}_1}-m^2_{\tilde{\chi}^{\pm}_2}}.
\eea

\begin{figure}
\begin{flushleft}
\mbox{\epsfxsize=16cm\epsffile{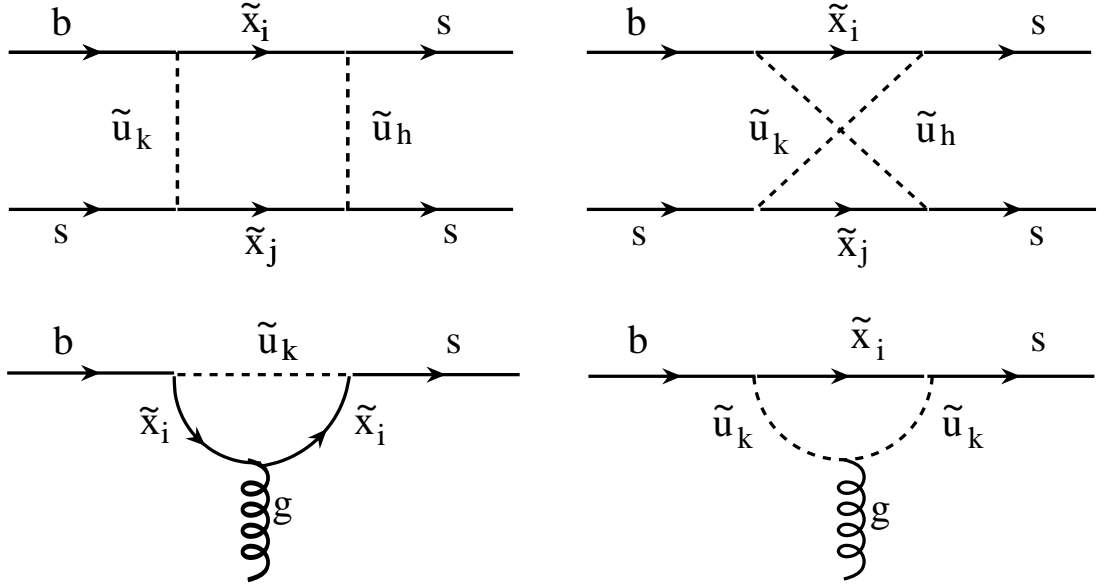}}
\caption{Box and penguin diagrams in $b \to s \bar s s$.}
\label{diam}
\end{flushleft}
\end{figure}

We calculated Wilson coefficients of SUSY diagrams in Fig.~\ref{diam} with chargino loops ~\cite{hm,loop1,loop2,ch}.

\begin{eqnarray}
\label{eqn3}
C_3^{SUSY} & = & \sum_{i,h,j,k} -\frac{\alpha_w^2}{4 m_{\tilde \chi_j}^2}
(G^{\ast iks}_{u_L} - H^{\ast iks}_{u_R})(G^{ihs}_{u_L} 
- H^{ihs}_{u_R})(G^{\ast jhs}_{u_L} - H^{\ast jhs}_{u_R})
(G^{jkb}_{u_L} - H^{jkb}_{u_R}) \nonumber \\
&&  G(x_{m_{{\tilde u}_k},m_{{\tilde x}_j}}, 
x_{m_{{\tilde u}_k},m_{{\tilde x}_j}},x_{m_{{\tilde u}_k},
m_{{\tilde x}_j}})  \nonumber \\ 
&& + \sum_{j,k} \frac{\alpha_s \alpha_w}{6 m_{{\tilde \chi}_j}^2} (G^{\ast jks}_{u_L} - H^{\ast jks}_{u_R}) (G^{\ast jkb}_{u_L} - H^{\ast jkb}_{u_R}) C_1(x_{m_{{\tilde u}_k},m_{{\tilde x}_j}}) \nonumber \\
&& + \sum_{j,k} \frac{2 \alpha_w \alpha_1}{9 m_{{\tilde \chi}_j}^2} (G^{\ast jks}_{u_L} - H^{\ast jks}_{u_R}) (G^{\ast jkb}_{u_L} - H^{\ast jkb}_{u_R})  \nonumber \\
&& \left(- C_1(x_{m_{{\tilde u}_k},m_{{\tilde x}_j}}) + C_2(x_{m_{{\tilde u}_k},m_{{\tilde x}_j}}) \right)  \nonumber \\
\label{eqn:C3}
C_4^{SUSY} &=& \sum_{j,k} -\frac{\alpha_s \alpha_w}{2 m_{{\tilde \chi}_j}^2} (G^{\ast jks}_{u_L} - H^{\ast jks}_{u_R}) (G^{\ast jkb}_{u_L} - H^{\ast jkb}_{u_R})  C_1(x_{m_{{\tilde u}_k},m_{{\tilde x}_j}}) \nonumber \\
\label{eqn:C4}
C_5^{SUSY} &=&  \sum_{i,h,j,k} \frac{\alpha_w^2}{2 m_{\tilde \chi_j}^2}
(G^{\ast iks}_{u_L} - H^{\ast iks}_{u_R}) H^{ihs}_{u_L} H^{\ast jhs}_{u_L}
(G^{jkb}_{u_L} - H^{jkb}_{u_R}) \nonumber \\
&& x_{m_{{\tilde \chi}_i},m_{{\tilde x}_j}} G_1(x_{m_{{\tilde u}_k},m_{{\tilde x}_j}}, x_{m_{{\tilde u}_k},m_{{\tilde x}_j}},x_{m_{{\tilde u}_k},m_{{\tilde x}_j}}) \nonumber \\ 
&& + \sum_{j,k} \frac{\alpha_s \alpha_w}{6 m_{{\tilde \chi}_j}^2} (G^{\ast jks}_{u_L} - H^{\ast jks}_{u_R}) (G^{\ast jkb}_{u_L} - H^{\ast jkb}_{u_R}) C_1(x_{m_{{\tilde u}_k},m_{{\tilde x}_j}})  \nonumber \\
&& + \sum_{j,k} \frac{2 \alpha_w \alpha_1}{9 m_{{\tilde \chi}_j}^2} (G^{\ast jks}_{u_L} - H^{\ast jks}_{u_R}) (G^{\ast jkb}_{u_L} - H^{\ast jkb}_{u_R}) \nonumber \\
&& \left( - C_1(x_{m_{{\tilde u}_k},m_{{\tilde x}_j}}) +  C_2(x_{m_{{\tilde u}_k},m_{{\tilde x}_j}}) \right) \nonumber \\
\label{eqn:C5}
C_6^{SUSY} &=&  \sum_{j,k} -\frac{\alpha_s \alpha_w}{2 m_{{\tilde \chi}_j}^2} (G^{\ast jks}_{u_L} - H^{\ast jks}_{u_R}) (G^{\ast jkb}_{u_L} - H^{\ast jkb}_{u_R}) C_1(x_{m_{{\tilde u}_k},m_{{\tilde x}_j}}) \nonumber \\
\label{eqn:C6}
C_g^{SUSY} &=&  \sum_{j,k} \frac{\alpha_w \pi}{2 m_{{\tilde \chi}_j}^2} (G^{\ast jks}_{u_L} - H^{\ast jks}_{u_R}) (G^{\ast jkb}_{u_L} - H^{\ast jkb}_{u_R}) D_1(x_{m_{{\tilde u}_k},m_{{\tilde x}_j}}) \nonumber \\
&& + \sum_{j,k} \frac{\alpha \pi}{2 m_{{\tilde \chi}_j}^2} (G^{\ast jks}_{u_L} - H^{\ast jks}_{u_R}) H^{\ast jkb}_{u_R}\frac{m_{{\tilde \chi}_j}}{m_b} D_2(x_{m_{{\tilde u}_k},m_{{\tilde x}_j}}) \\
\label{eqn:Cg}
\end{eqnarray}
where
\be
\label{eqn:coe}
G_{u_L}^{jki} = V_{j1}^* (\Gamma_{u_L} V_{CKM})^{ki}, \ \ \ \ \ \ H_{U_R}^{jki} = V_{j2}^*(\Gamma_{U_R} V_{CKM})^{ki} \frac{M_U}{\sqrt{2} M_w sin\beta}
\ee

$m_{{\tilde \chi}_j}$, $j=1,2$ are chargino masses, and  $x_{m_A, m_B} = m_A^2/m_B^2$. The loop functions are given in Appendix. 

In addition to Wilson coefficients, the hadronic matrix elements of the operators are also needed in computing the SUSY contributions to the CP asymmetry in $B \to \phi K_S$ decay. We use the results in ref. \cite{hm,hadron} 
\begin{eqnarray}
\langle \phi K_S|{\Oi}_{3,4}|\bar{B}^0_d\rangle &=& \frac{1}{4}
H \left( 1+\frac{1}{N_c}\right) \nonumber \\
\langle \phi K_S|{\Oi}_5|\bar{B}^0_d\rangle &=& \frac{1}{4} H \nonumber \\
\langle \phi K_S|{\Oi}_6|\bar{B}^0_d\rangle &=& \frac{1}{4} H
\frac{1}{N_c} \nonumber \\
\langle \phi K_S|{\Oi}_{g}|\bar{B}^0_d\rangle &=& \kappa
\frac{\alpha_s}{2\pi} H \frac{N_c^2-1}{2N_c^2}.\\
\end{eqnarray}
where the analytical expression for $H$ is in 
Appendix B of ref.\cite{hm}. 
The value of $\kappa$ is  approximately -1.1 ~\cite{susy2,hm}.

\section{Numerical results}\label{num}
With the formulae in section~\ref{cp1}, we can numerically analyze the chargino contributions in B system.  There are three kinds of mixing in the up-squark mass matrix: Left-Left, Right-Right and Left-Right. 
Each mixing contributes a mixing angle and a phase angle in 
$S_{\phi K_S}$ computation. However, the phase of $\mu$ is 
strongly constrained by the neutron electric dipole moment (EDM). 
We have checked that our results have no substantially dependence 
on $\mu$. To simplify our computation, we take the phase of $\mu$ to be zero in this paper. 

The agreement between the CP asymmetry in $B \to J/\psi K_S$ decay 
observed by BaBar and Belle with the SM prediction leaves very little room 
for new physics contributions in $S_{J/\psi K_S}$. 
It strongly constrains the SUSY contributions to  $S_{J/\psi K_S}$ 
and strongly  suppresses the CP violating phase in $B \to \phi K_S$ system ~\cite{constaint}, which will be noticed in the following analysis. 

In order to simplify the analysis, we consider three special cases. 
First, left-left mixing dominates, LL scenario. 
Right-right mixing  is very small and can be ignored in this scenario.  
The second scenario is right-right mixing dominant, RR scenario, while the 
last one has comparable left-left and right-right mixing contributions, 
LL + RR scenario. 
We need to consider both contributions in the last case. 
Because of the large parameter space, we only show some results for 
RR + LL scenario with specific simplifications.
\subsection{Left-Left Mixing Dominant}
In this scenario, left-left mixing is the only 
sizable mixing and there is only one phase and one mixing angle. 
We take the mixing matrix to be: 
\begin{equation}
\label{llmatrix}
R = \left(
\begin{array}{cc}
 cos\theta_{23} & sin\theta_{23} e^{i\phi_L} \\
 -sin\theta_{23}  & cos\theta_{23} e^{-i\phi_L}
\end{array} \right).
\end{equation}
where $\theta_{23}$ is the mixing angle. 
The subscript $23$ represents the left-left mixing and is the mixing
between second and third rows in up-squark mass matrix. 
$\phi_L$ is the phase angle. 
We expect that these mixing and phase angles can significantly alter the 
value of $S_{\phi K}$. We have checked that the precise values of $M_2$ 
and $\mu$ do not have 
a substantial effect on our results. 
We use a range of values of $M_2$ and $\mu$ from 2 GeV to about 400 GeV. 
We obtain almost the same results. 
Thus, we take $M_2$ and $\mu$ to be 200 GeV in most of our calculations. 

In this scenario, we have parameters $$ m_{{\tilde L}_2}, m_{{\tilde L}_3},  m_{{\tilde \chi}_2},  m_{{\tilde \chi}_1}, tan\beta, \theta_{23}, \phi_L .$$ with $ m_{{\tilde L}_3} <  m_{{\tilde L}_2}$ and $ m_{{\tilde \chi}_1} <  m_{{\tilde \chi}_2}$. 
As pointed in ref.~\cite{constaint}, the imaginary parts of this matrix are 
stringently constrained by $B \to J/\psi K_S$ decay. 
In order for the mixing angle $\theta_{23}$ to be less constrained, the 
phase $\phi_L$ we use in our computation is very severe constrained, 
$\phi_L \leq 10^{-2}$.

\begin{figure}
\begin{center}
\mbox{\epsfxsize=18cm\epsffile{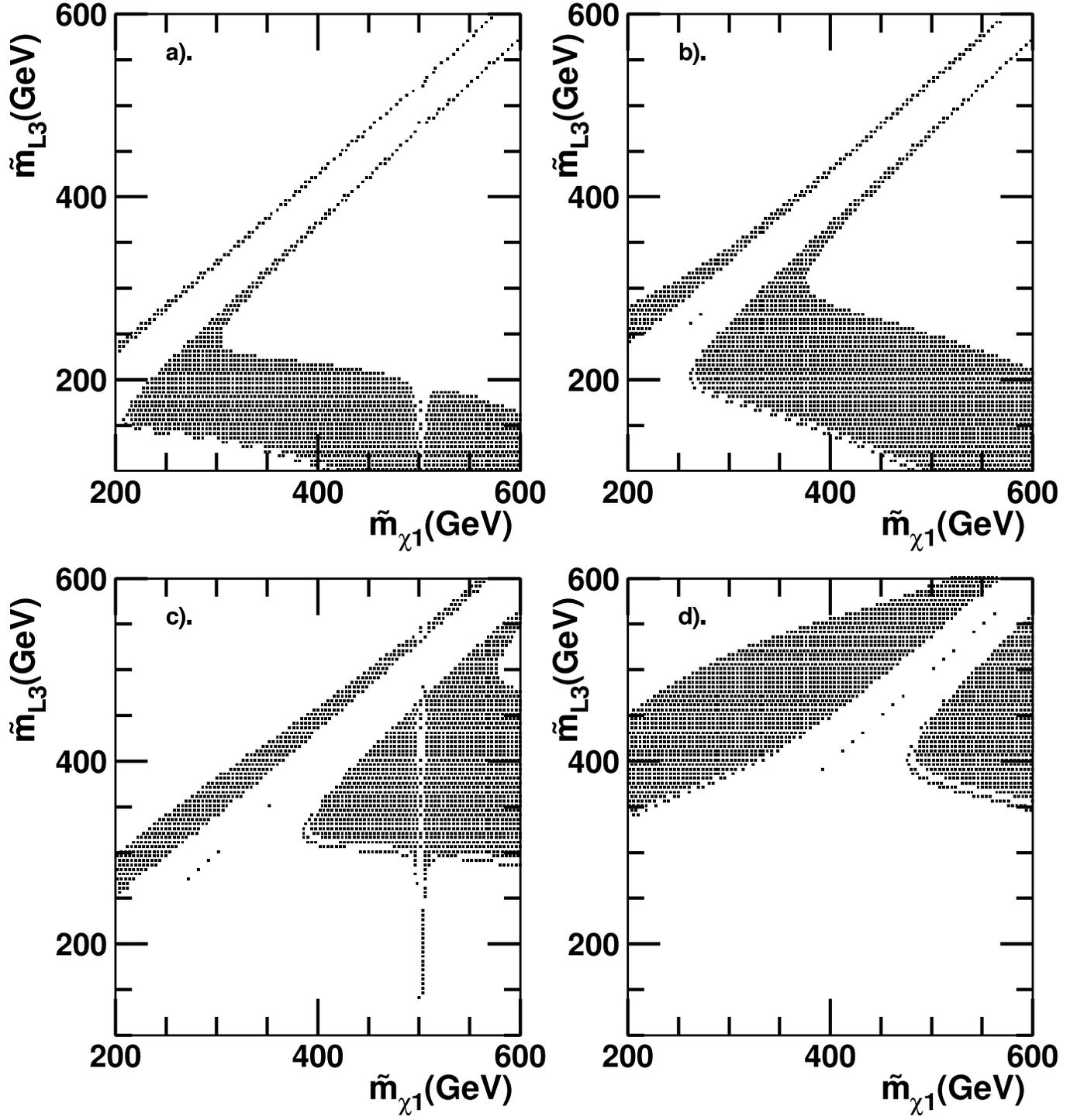}}
\vspace{3mm}
\caption[]{In left-left mixing scenario, $S_{\phi K_S}$ in the plane of $\tilde{m}_{L3}$ vs $\tilde{m}_{\chi_1}$. The shaded region corresponds to $ -0.7 < S_{\phi K_S} < 0.7$ with $\phi = 0.01$ and $\theta = \pi/4$.  (a). $\kappa = -1.1$, $\tilde{m}_{\chi 2} = 10 \tilde{m}_{\chi 1}$; (b). $\kappa = -1.1$, $\tilde{m}_{\chi 2} = 100 \tilde{m}_{\chi 1}$; (c).  $\kappa = -2.$, $\tilde{m}_{\chi 2} = 10 \tilde{m}_{\chi 1}$; (d). $\kappa = -2.$, $\tilde{m}_{\chi 2} = 100 \tilde{m}_{\chi 1}$;  }
\label{ll:chi}
\end{center}
\end{figure}

There are seven parameters. Their effects on $S_{\phi K_S}$ are different. 
In Fig.~\ref{ll:chi}, we plot $S_{\phi K}$ as a 
function of $ m_{{\tilde L}_3}$ and $ m_{{\tilde \chi}_1}$ within the range 
$-0.7 < S_{\phi K} < 0.7 $.  We fixed $tan\beta = 5$, and $ m_{{\tilde L}_2} = 5$~TeV. 
$\theta_{23}$ is taken to be $\pi/4$, and $\phi_L = 0.01$. 
In frame (a), The value of $\kappa $ is equal to  -1.1 and $ m_{{\tilde \chi}_2} = 10  m_{{\tilde \chi}_1}$. In frame (b), $\kappa$ has the same value as in frame (a), but $ m_{{\tilde \chi}_2} = 100  m_{{\tilde \chi}_1}$. We change $\kappa$ to -2 in frame (c) and  frame (d) with $ m_{{\tilde \chi}_2} = 10  m_{{\tilde \chi}_2}$ in frame (c) and  $ m_{{\tilde \chi}_2} = 100  m_{{\tilde \chi}_2}$ in frame (d). 
Within the shaded region, $ S_{\phi K}$ is larger than -0.7, 
but smaller than 0.7. From this graph, we notice that:
\begin{itemize}
\item{ There are many assumptions in the process of $\kappa$ estimation. 
$\kappa = -1.1$ is just an approximate value. We present results for 
different values of $\kappa$ to show the dependence of 
$S_{\phi K_S}$ on $\kappa$.  
The main SUSY contributions in this scenario come from penguin diagrams. 
The hadronic matrix of the corresponding operator is proportional to $\kappa$ 
and $\kappa$ surely has some effect on $S_{\phi K_S}$. 
This can be observed through the difference among graphs in 
Fig.~\ref{ll:chi}.}
\item{ We show how large the effect of the parameter $\kappa$ is. 
We take $\kappa$ to be -2 in frame (c) and frame (d). 
Except for these two frames, $\kappa $  is fixed to -1.1 in all other graphs.}
\item { We focus on chargino loop contributions. 
Any changes in the chargino parameters have a large influence on our results. 
As is seen in frame (c) and frame (d),  increasing $m_{{\tilde \chi}_2}$ 
changes the allowed values significantly. }
\item{ There are big gaps around the line 
$m_{{\tilde L}_3} = m_{{\tilde \chi}_1}$. 
We refer the reader to the loop functions in Eq.~\ref{eqn:Cg}, 
which are presented explicitly in Appendix. When the chargino and squark 
masses are equal, the functions in the Appendix take a different 
form and cause the big gaps in the graphs.} 
\item{ The small discontinuity in 
frame (a) and frame (b) at $m_{{\tilde \chi}_1} = 500$~GeV arises for the 
same reason. That is because $m_{{\tilde \chi}_2}$ is ten times bigger than $m_{{\tilde \chi}_1}$, and at the point of $m_{{\tilde \chi}_1} = 500$~GeV, $m_{{\tilde \chi}_2} = 5$~TeV, which is equal to $m_{{\tilde L}_2}$. }   
\end{itemize}

We further explore the dependence of $S_{\phi K_S}$ on the phase angle $\phi_L$
and the mixing angle $\theta_{23}$ in Fig.~\ref{ll:phi} and Fig.~\ref{ll:theta}. We reduce the value of $\phi_L$ to 0.005 in Fig.~\ref{ll:phi}. 
We continue to take 
tan$\beta$ = 5, $m_{{\tilde L}_2} = 5$~TeV and $m_{{\tilde \chi}_2} = 10 m_{{\tilde \chi}_1}$. 
As shown in the plot, with a smaller $\phi_L$ angle the allowed region is 
suppressed $m_{{\tilde L}_3} < 200$~GeV. 
In Fig.~\ref{ll:theta}, $\theta_{23}$ is reduced to $\pi/6$ and 
$\phi_L$ is fixed to be 0.01. The allowed region changes as a result.
Fig.~\ref{ll:theta} presents an even bigger change as $\theta$ decreases 
from $\pi/4$ to $\pi/6$.
 
There is one parameter left: tan$\beta$, which is very important. 
From Eq.~\ref{eqn:mixing} and Eq.~\ref{eqn:coe}, we can see that tan$\beta$ 
affects not only the mixing angles, but also the Wilson Coefficients. 
Any change in tan$\beta$ alters $S_{\phi K_S}$ significantly.  
We plot Fig.~\ref{ll:tan} to explore the dependence of $S_{\phi K_S}$ on 
tan$\beta$. Apart from tan$\beta$, all the other parameters remain the same 
as the parameters in frame (a) of Fig.~\ref{ll:chi}. In frame (a), we take 
tan$\beta$ to be 20, while in frame (b), tan$\beta$ is set to be 40.  
Several features are worth noting:
\begin{itemize}
\item{ By comparing two graphs in this figure, we find that the 
allowed 
region permits larger masses with the increase in tan$\beta$.}
\item{ As for graph in frame (b), we just plot $S_{\phi K_S}$ in a region 
with the masses of $m_{{\tilde L}_3}$ and  $m_{{\tilde \chi}_1}$ 
smaller than 1.5 TeV. 
There is a large allowed region beyond 1.5 TeV, which does not appear 
in this graph. Still, we can see the strong dependence of $S_{\phi K_S}$ on
tan$\beta$.
}
\end{itemize}

\subsection{Right-Right Mixing Dominant}

This corresponds to the case where left-left mixing is very small 
and can be ignored. The dominant mixing is right-right mixing. the mixing matrix is:
\begin{equation}
\label{rrmatrix}
R = \left(
\begin{array}{cc}
 cos\theta_{56} & sin\theta_{56} e^{i\phi_R} \\
 -sin\theta_{56}  & cos\theta_{56} e^{-i\phi_R}
\end{array} \right).
\end{equation}
The free parameter are now 
$$ m_{{\tilde R}_2}, m_{{\tilde R}_3},  m_{{\tilde \chi}_2},  m_{{\tilde \chi}_1}, tan\beta, \theta_{56}, \phi_R.$$ with the same constraints  $ m_{{\tilde R}_3} <  m_{{\tilde R}_2}$ and $ m_{{\tilde \chi}_1} <  m_{{\tilde \chi}_2}$.

Unlike the Left-Left dominant scenario, this time, we fix 
$m_{{\tilde R}_3} = 300$~GeV. 
In Fig.~\ref{rr:chi}, we plot $S_{\phi K_S}$ as a function of $ m_{{\tilde R}_2}$ and $ m_{{\tilde \chi}_1}$. tan$\beta$  is still 5.  
We try different values of $\phi$ within the range $\phi \leq 0.01$, 
and find that in almost the whole range, the value of $S_{\phi K_S}$ 
is approximately 0.73, which is the same as the SM prediction. 
SUSY contributions are very small. We relax the $\phi$ constraint and 
take $\phi$ to be $\pi/4$.  $\theta_{56}$ is also $\pi/4$. 
In frame (a), $m_{{\tilde \chi}_2} = 10 m_{{\tilde \chi}_1}$, while in frame (b), $m_{{\tilde \chi}_2}$ is 20 times bigger than $m_{{\tilde \chi}_1}$. 
From this graph, we notice that, in contrast to the left-left mixing case, the shaded region in frame (b) is concentrated
toward small masses side than that in frame (a) 
. As shown in Eq.~\ref{eqn:mixing}, 
the difference between the chargino masses, 
$m_{{\tilde \chi}_2} - m_{{\tilde \chi}_1}$ affects the mixing angles in 
the chargino sector. 
The mixing angles determine the values of $S_{\phi K_S}$. 
We allow $m_{{\tilde \chi}_2}$ to be proportional to $m_{{\tilde \chi}_1}$ 
such that the larger the mass of  $m_{{\tilde \chi}_1}$, 
the bigger the mass difference of  $m_{{\tilde \chi}_2}$ 
and  $m_{{\tilde \chi}_1}$. 
It is the fact that $m_{{\tilde \chi}_2}$ in frame (b) is ten times bigger than the corresponding $m_{{\tilde \chi}_2}$ in frame (a) make the allowed $S_{\phi K_S}$ region in frame (a) concentrate on larger  $m_{{\tilde \chi}_1}$ , while in frame (b), $m_{{\tilde \chi}_1}$ is relatively small.  

At $m_{{\tilde \chi}_1} = 300$~GeV, we observe a small peak in the graphs. 
As in the left-left mixing case, at this point $m_{{\tilde \chi}_1}$ equals 
to $m_{{\tilde R}_3}$ , which uses a difference loop function.    

We also wish to explore the dependence of $S_{\phi K_S}$ on other parameters. 
Unlike in left-left mixing case, this time instead of studying the dependence of $S_{\phi K_S}$ on each parameter, 
we show a graph scanning the whole parameter  space: $$ m_{{\tilde L}_2}, m_{{\tilde L}_3}, m_{{\tilde \chi}_1}, m_{{\tilde \chi}_2}, \theta_{23}, \phi_R .$$
Within the constraints $m_{{\tilde R}_2} > m_{{\tilde R}_3}$ and $m_{{\tilde \chi}_2} > m_{{\tilde \chi}_1}$, we take:
\begin{eqnarray}
\label{eqn:params}
300 ~GeV & < m_{\tilde R2} < & 2500 ~GeV  \\ 
250 ~GeV & < m_{\tilde R3} < & 1000 ~GeV \nonumber  \\
100 ~GeV & < m_{{\tilde \chi}_1} < & 500 ~GeV \nonumber  \\
150 ~GeV & < m_{{\tilde \chi}_2} < & 1000 ~GeV \nonumber  \\ 
 0 & < \theta_{23} < & \pi/2 \nonumber\\ 
0 & < \phi_R < & 2\pi \nonumber 
\end{eqnarray}

The results are present in Fig.~\ref{rr:scan} with $S_{\phi K_S}$ vs $m_{{\tilde \chi}_2}$. 
In frame (a), the tan$\beta$ is 15, and in frame (b) it is increased to 50. 
In frame (a), we notice that many of the points are around 0.73, 
the value of standard model prediction. 
This behavior tells us with the parameters in Eq.~\ref{eqn:params} SUSY contributions are 
usually small and cannot alter $S_{\phi K_S}$ substantially.  What we are interested in 
is the regions of parameter space where SUSY contributions are dominant. 
In these regions, the value of $S_{\phi K_S}$ can be as low as negative -0.98. 
As tan$\beta$ increases, SUSY contributions are increased.  
As shown in frame (b), when tan$\beta$ is increased to 50, more points on the
scatter plot deviate from the SM prediction. 
The mixing angles $\theta_U$ and $\theta_V$ in the chargino sector vary 
with tan$\beta$ and  increase the SUSY contributions to $S_{\phi K_S}$. 
There are very few points within $m_{{\tilde \chi}_2} < 300$~GeV. 
That is because we require $m_{{\tilde \chi}_2} > m_{{\tilde \chi}_1}$, which 
limits the probability of small $m_{{\tilde \chi}_2}$.    

\subsection{Left-Left Mixing + Right-Right Mixing}

Left-left mixing and right-right mixing are comparable in this scenario. 
They both make contributions to $S_{\phi K_S}$. 
There are two mixing angles, two phase angles and four mass parameters. 
To simplify the computation, we assume that left and right squarks have the 
same masses in same generation, that is: $m_{{\tilde L}_2} = m_{{\tilde R}_2} = \tilde m_2$ and $m_{{\tilde L}_3} = m_{{\tilde R}_3} = \tilde m_3$. 
With these assumptions, the mixing angles 
$\theta_L$ in the left-left mixing and 
$\theta_R$ in the right-right mixing are the same, 
$\theta_L$ =  $\theta_R$ =  $\theta$ and also the phase angle, 
$\phi_L$ = $\phi_R$ = $\phi$. 
Therefore, the free parameters in this scenario are reduced to:  $$ \tilde m_2, \tilde m_3,  m_{{\tilde \chi}_2},  m_{{\tilde \chi}_1}, tan\beta, \theta, \phi.$$ with the same constraints  $\tilde m_3 < \tilde  m_2$ and $ m_{{\tilde \chi}_1} <  m_{{\tilde \chi}_2}$. 

We display our main results in a few plots. In Fig.~\ref{lr:chi}, we fixed  $\tilde m_2 = 5$~TeV and the mixing angle $\theta$ is set to be $\pi/4$. 
The phase angle $\phi$ is 0.01. 
In frame (a), we use $m_{{\tilde \chi}_2} = 10 m_{{\tilde \chi}_1}$ , 
while in frame (b)  $m_{{\tilde \chi}_2}$ is 50 times bigger than  
$m_{{\tilde \chi}_1}$. This graph is very similar to the corresponding graph 
in left-left mixing scenario, Fig~\ref{ll:chi}, except here we have a larger 
allowed parameter space. As in the LL dominant scenario, 
chargino masses affect $S_{\phi K_S}$ significantly. This is shown in the 
difference between graphs in frame (a) and  frame (b). 
As expected, left-left mixing provides the dominant contribution and 
right-right mixing adds to the allowed region. 
The big gaps around line ${\tilde m}_3 = m_{{\tilde \chi}_1}$ and the small 
gap at ${\tilde m}_2 = m_{{\tilde \chi}_2}$ are much more obvious than in 
left-left dominant scenario. We can conclude that left-left plus right-right mixing together make a big contribution to $S_{\phi K_S}$.

$S_{\phi K_S}$ also depends on  the mixing and phase angles.  
Fig.~\ref{lr:phi} shows the effect of phase angle $\phi$, 
while Fig.~\ref{lr:theta} shows the effect of the mixing angle $\theta$. 
We keep $\tilde{m}_2 = 5$~TeV and 
$m_{{\tilde \chi}_2} = 10 m_{{\tilde \chi}_1}$. In Fig.~\ref{lr:phi}, the phase angle $\phi$ is reduced to $0.005$ and mixing angle is fixed to be $\pi/4$, 
while in Fig.~\ref{lr:theta}, 
$\theta $ changes to $ \pi/6$ and $\phi$ is kept at 0.01. 
As in the LL case, 
as $\phi$ decreases the allowed parameter space decreases and as 
$\theta$ becomes smaller the allowed masses also become 
smaller.

Fig.~\ref{lr:scan} is the scanned graph. We scan the same parameter space as 
in the right-right mixing case. We find that in frame (a), although there is 
still a concentration of points around the standard model value, 
more points  deviate from their SM prediction than with only 
right-right mixing. This effect is even more obvious in frame (b). 
This also shows a stronger contribution in this scenario.
  
\section{Conclusion}\label{con}
We have studied the supersymmetric contributions, especially the 
chargino loop contributions, to CP asymmetry in $B \to \phi K_S$ decay. 
To emphasize chargino loop contributions, we work in a special basis where 
the down-type squark mass matrix is diagonal. 
In this special basis, gluino loop contributions, which are very important in 
most SUSY frames, are ruled out. 
Chargino loops are the main contributions to $S_{\phi K_S}$. 

Within the allowed parameter region, we studied three special cases. 
Left-left mixing dominant, right-right mixing dominant and 
left-left plus right-right mixing dominant. In the LL case, 
chargino contributions to the CP asymmetry can be much larger than the SM 
contributions in some regions of parameter space. 
The CP asymmetry $S_{\phi K_S}$ varies significantly as the chargino masses 
vary. In the RR case, things are very similar with the LL case, except 
the allowed region is totally different from the  LL case. 
The LL plus RR mixing gives the largest contribution to $S_{\phi K_S}$. 

We consider SUSY contributions to CP asymmetry $S_{\phi K_S}$ with only 
chargino loops. In most SUSY scenarios, gluino loops will provide very 
large contributions, which may be larger than chargino contributions for some
parameters.
Gluino, chargino and even neutralino loops make up the whole SUSY 
contribution. 

$b \to s\gamma$ decay puts  very strong limits on the allowed parameter space. 
It excludes some regions in our graphs. 
The calculation of $b \to s \gamma$  is very model dependent. 
We are not including the $b \to s \gamma$ constraints in our paper. 
But we can still conclude that SUSY is a candidate to explain the 
CP asymmetry deviation in $B \to \phi K_S$ and $B \to J/\psi K_S$ 
and the contributions of chargino loops play an important role 
in some regions of SUSY parameter space.

\section*{Acknowledgments}
It is a pleasure to thank Professor G. Valencia for his helpful comments and 
valuable information in the loop function  computation. Thanks Professor D. Atwood for the discussion on this paper. Thanks Dr. J. Abraham to help me editing this paper. This research was supported in part by the U.S. Department of Energy
under contracts number DE-FG02-01ER41155
\appendix
\section{ Loop Functions}
\begin{eqnarray}
    G(A,B,C) &=&  \int_0^\infty \frac{ z^2 \, dz}{(z+1)
                (z+A) (z+B)(z+C)} \nonumber \\ 
        &=& 
    \frac{A^2 \log A}{(A-1)(A-B)(A-C)}
    +\frac{B^2 \log B}{(B-1)(B-A)(B-C)}
    \nonumber \\ &&
    +\frac{C^2 \log C}{(C-1)(C-A)(C-B)},
    \\
    \nonumber  \\
    G1(A,B,C) &=&  \int_0^\infty \frac{ z \, dz}{(z+1)
                (z+A) (z+B)(z+C)} \nonumber \\ 
        &=& 
    -\frac{A\log A}{(A-1)(A-B)(A-C)}
    -\frac{B \log B}{(B-1)(B-A)(B-C)}
    \nonumber \\ &&
    -\frac{C \log C}{(C-1)(C-A)(C-B)},
\end{eqnarray}

\begin{eqnarray}
    C_1 (x) &=&
    \frac{2x^3-9x^2+18x-11-6\log x}{36(1-x)^4},
    \\
    \nonumber \\
    C_2 (x) &=&
    \frac{-16x^3+45x^2-36x+7+6x^2(2x-3)\log x}{36(1-x)^4},
    \\
    \nonumber  \\
    D_1 (x) &=&
    \frac{-x^3+6x^2-3x-2-6x\log x}{6(1-x)^4},
    \\
     \nonumber \\
    D_2 (x) &=&
    \frac{-x^2+1+2x\log x}{(x-1)^3},
\end{eqnarray}


%
\iftightenlines\else\newpage\fi
\iftightenlines\global\firstfigfalse\fi
\def\dofig#1#2{\epsfxsize=#1\centerline{\epsfbox{#2}}}
\def\fig#1#2{\epsfxsize=#1{\epsfbox{#2}}}
%
\newpage

\newpage

\newpage
\begin{figure}
\noindent
\dofig{15cm}{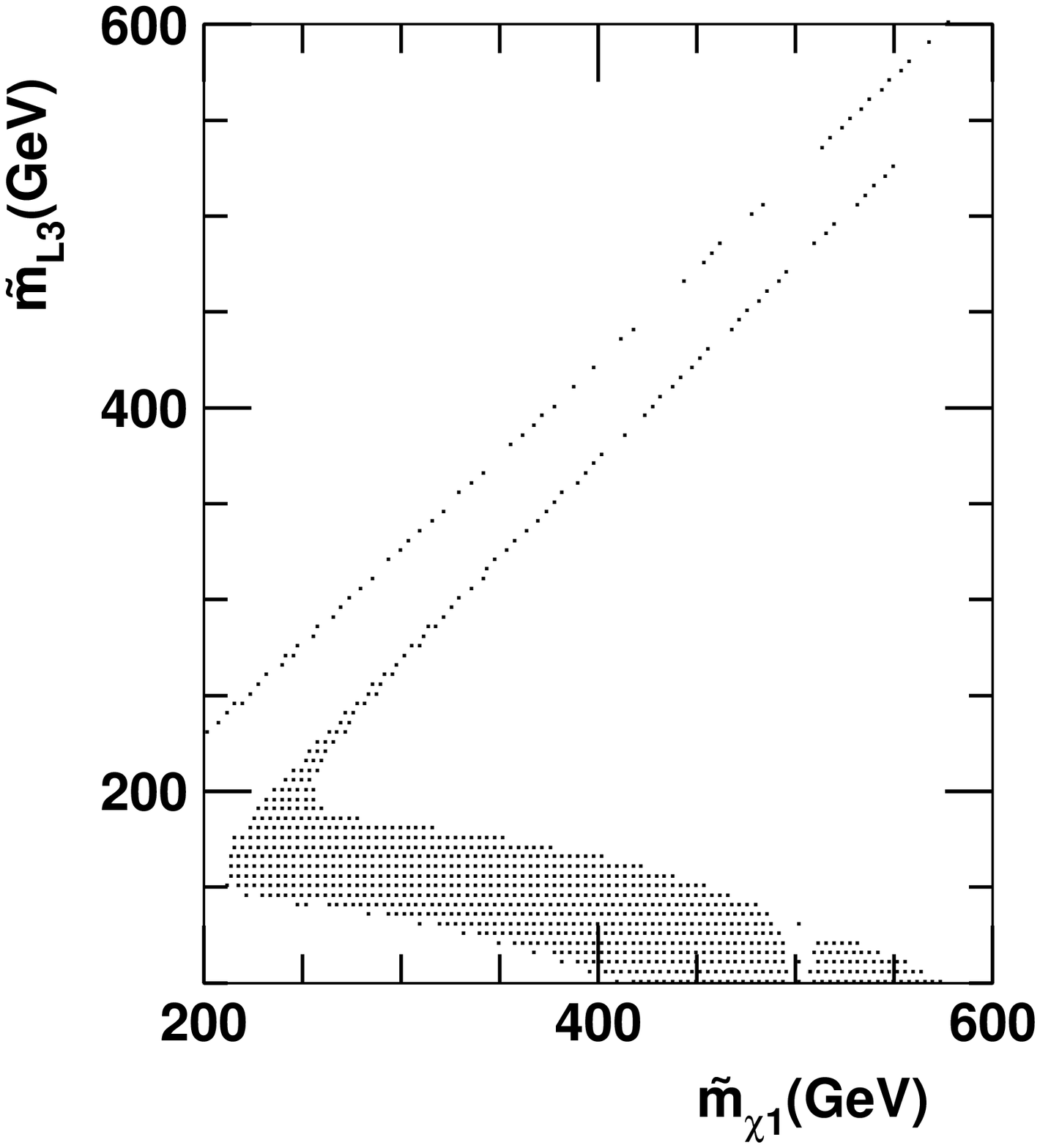}
\caption{In left-left mixing scenario, $S_{\phi K_S}$ in the plane of $\tilde{m}_{L3}$ vs $\tilde{m}_{\chi_1}$. The shaded region corresponds to $ -0.7 < S_{\phi K_S} < 0.7$ with $\phi = 0.005$, $\theta = \pi/4$ and $\kappa = -1.1$. $\tilde{m}_{\chi 2} = 10 \tilde{m}_{\chi 1}$
}
\label{ll:phi}
\end{figure}

\newpage

\begin{figure}
\dofig{15cm}{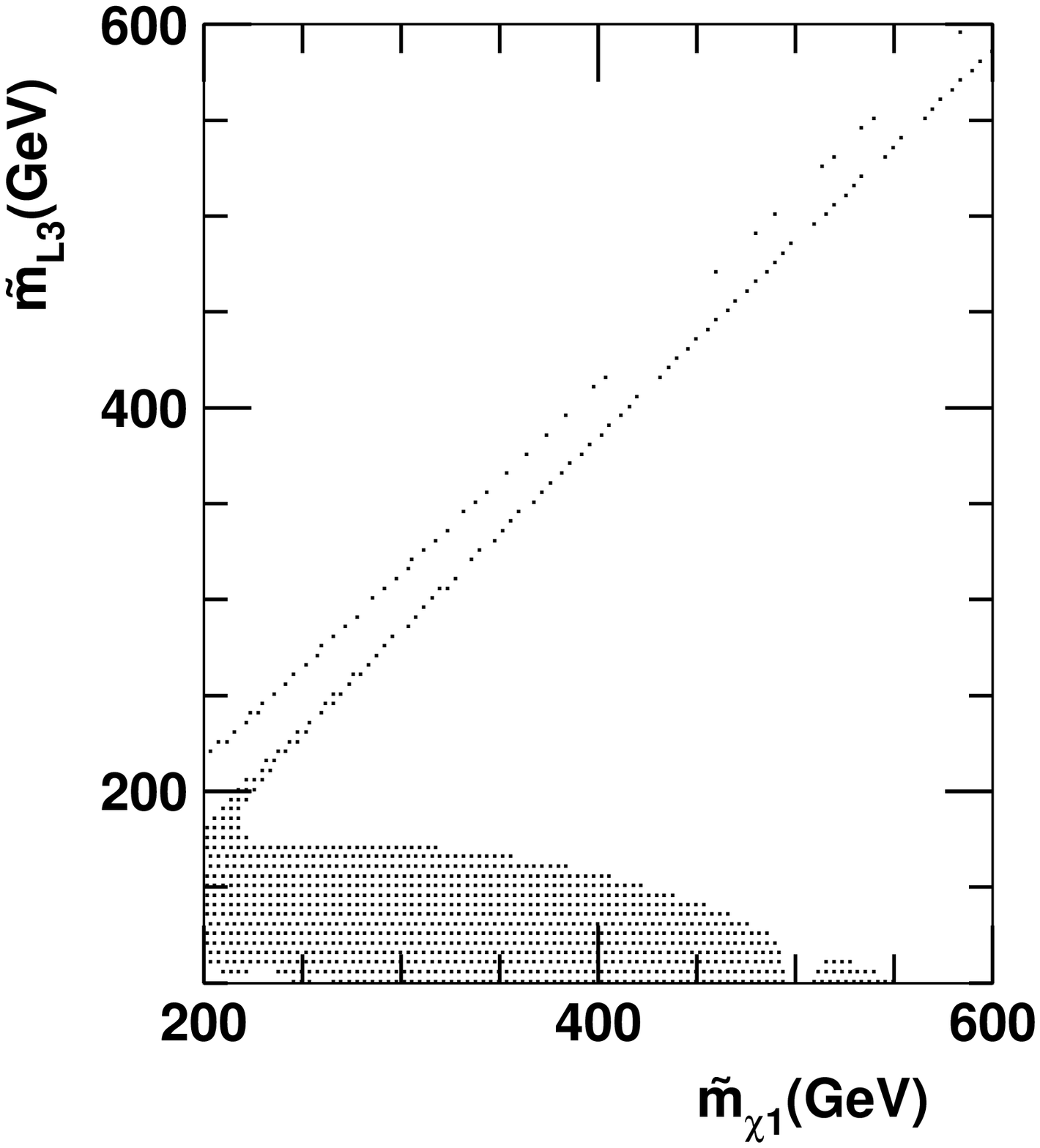}
\caption{In left-left mixing scenario, $S_{\phi K_S}$ in the plane of $\tilde{m}_{L3}$ vs $\tilde{m}_{\chi_1}$. The shaded region corresponds to $ -0.7 < S_{\phi K_S} < 0.7$ with $\phi = 0.01$, $\theta = \pi/6$ and $\kappa = -1.1$. $\tilde{m}_{\chi 2} = 10 \tilde{m}_{\chi 1}$
}
\label{ll:theta}
\end{figure}
\newpage
\begin{figure}
\dofig{18cm}{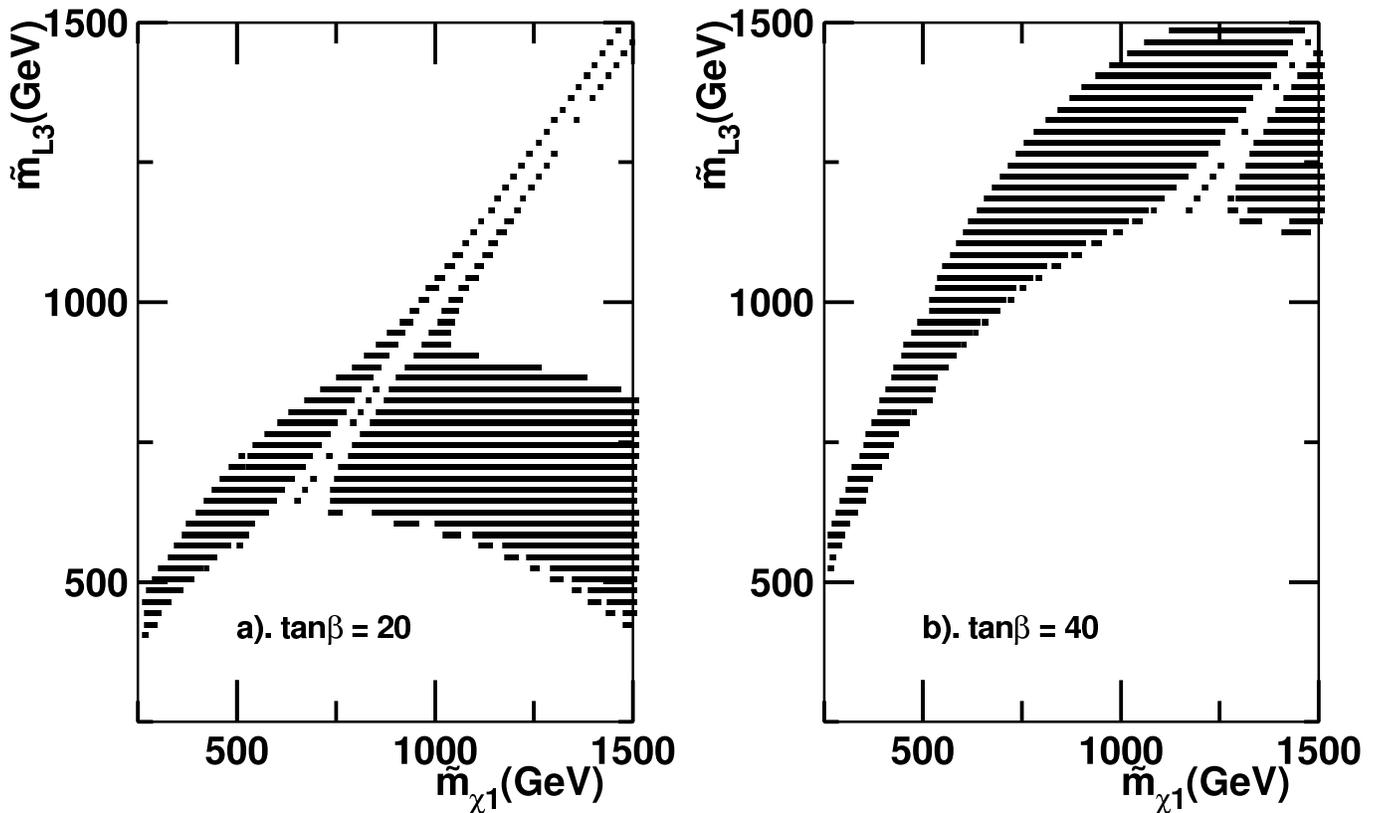}
\caption[]{In left-left mixing scenario, $S_{\phi K_S}$ in the plane of $\tilde{m}_{L3}$ vs $\tilde{m}_{\chi_1}$. The parallel lined region corresponds to $ -0.7 < S_{\phi K_S} < 0.7$ with $\phi = 0.01$, $\theta = \pi/4$ and $\kappa = -1.1$. $\tilde{m}_{\chi 2} = 10 \tilde{m}_{\chi 1}$. (a) tan$\beta$ = 20; (b) tan$\beta$ = 40.}
\label{ll:tan}
\end{figure}
\newpage

%
\begin{figure}
\dofig{16cm}{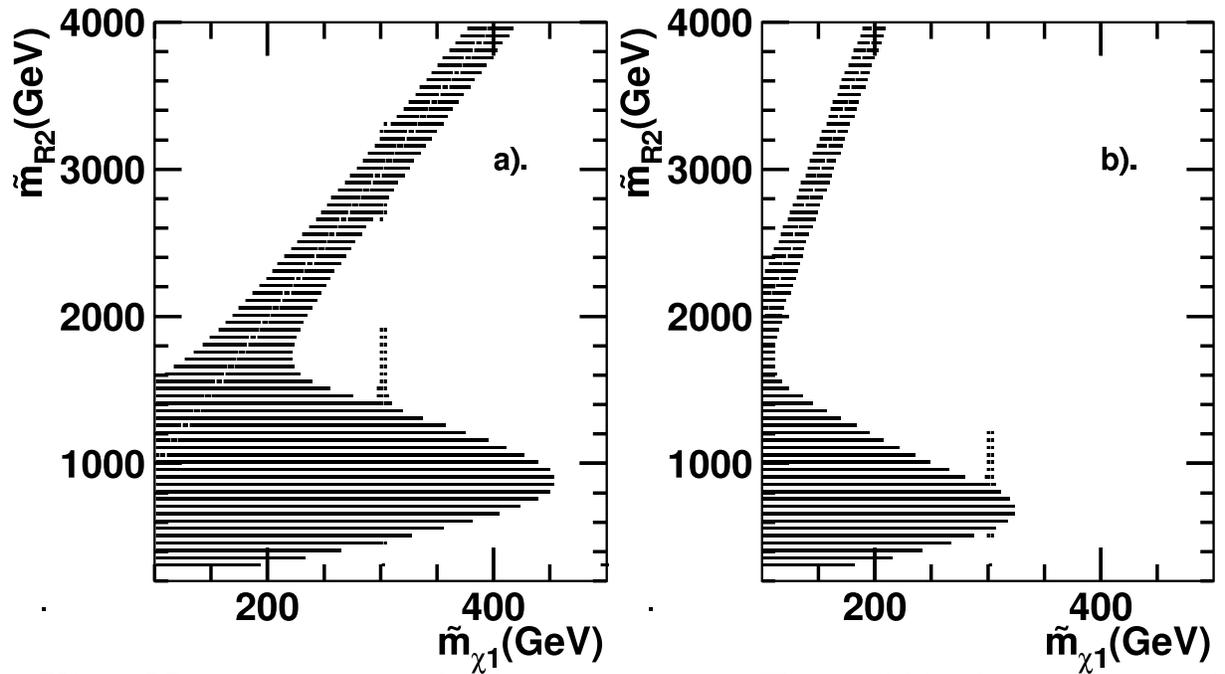}
\caption{RR scenario. $S_{\phi K_S}$ in the plane of $\tilde{m}_{L3}$ vs $\tilde{m}_{\chi_1}$. The parallel lined region corresponds to $ -0.7 < S_{\phi K_S} < 0.7$ with $\phi = \pi/4$, $\theta = \pi/4$ and $\kappa = -1.1$. (a) $\tilde{m}_{\chi 2} = 10 \tilde{m}_{\chi 1}$; (b) $\tilde{m}_{\chi 2} = 100 \tilde{m}_{\chi 1}$; 
}
\label{rr:chi}
\end{figure}
\newpage

\begin{figure}
\dofig{16cm}{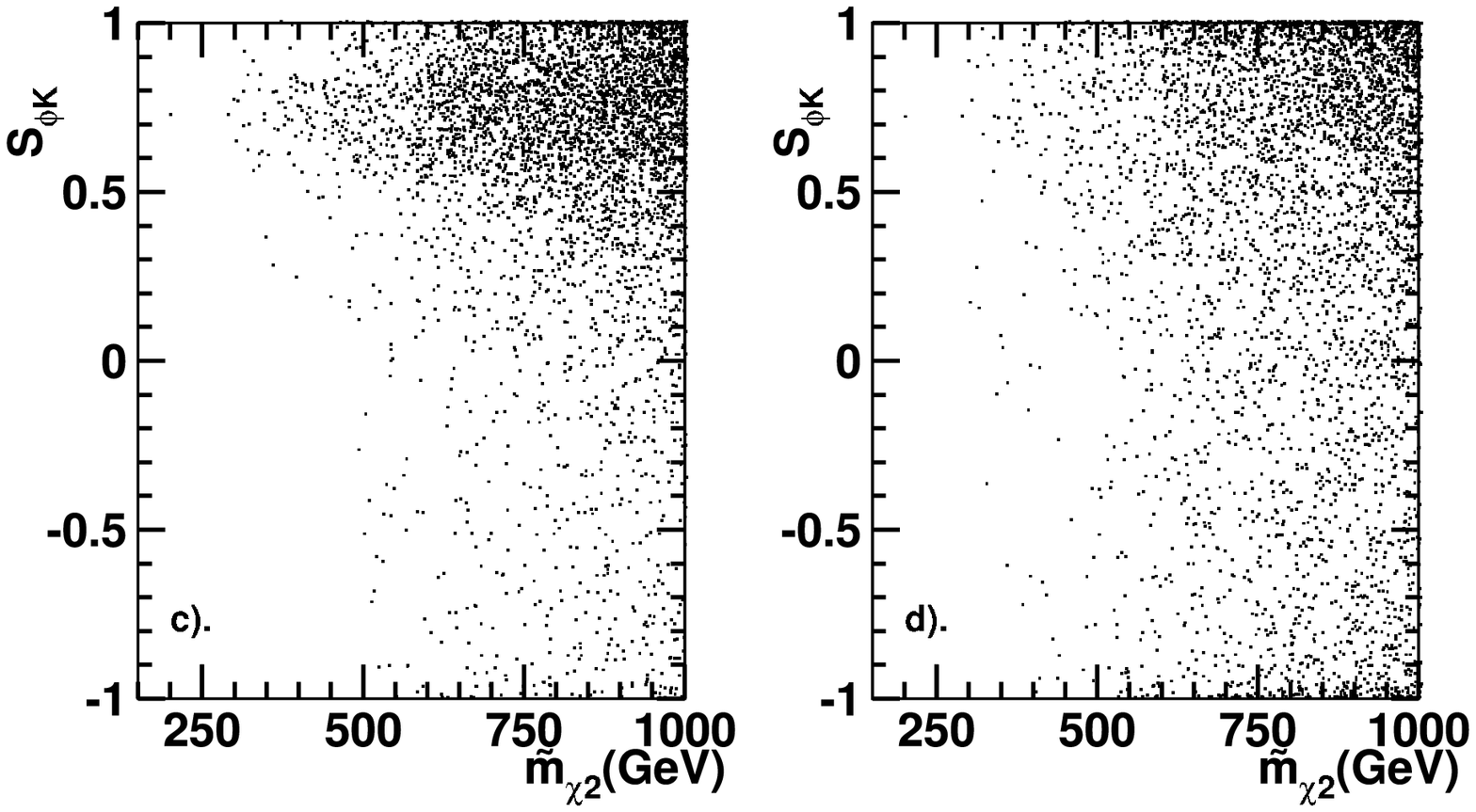}
\caption{ RR case. The CP asymmetry $S_{\phi K_S}$ scanned within parameter space over the region mentioned in the text. We show the results in the plane of  $S_{\phi K_S}$ vs $\tilde{m}_{L3}$ with (a). tan$\beta$ = 15 and (b). tan$\beta$ = 50}
\label{rr:scan}
\end{figure}
\newpage
\begin{figure}
\dofig{16cm}{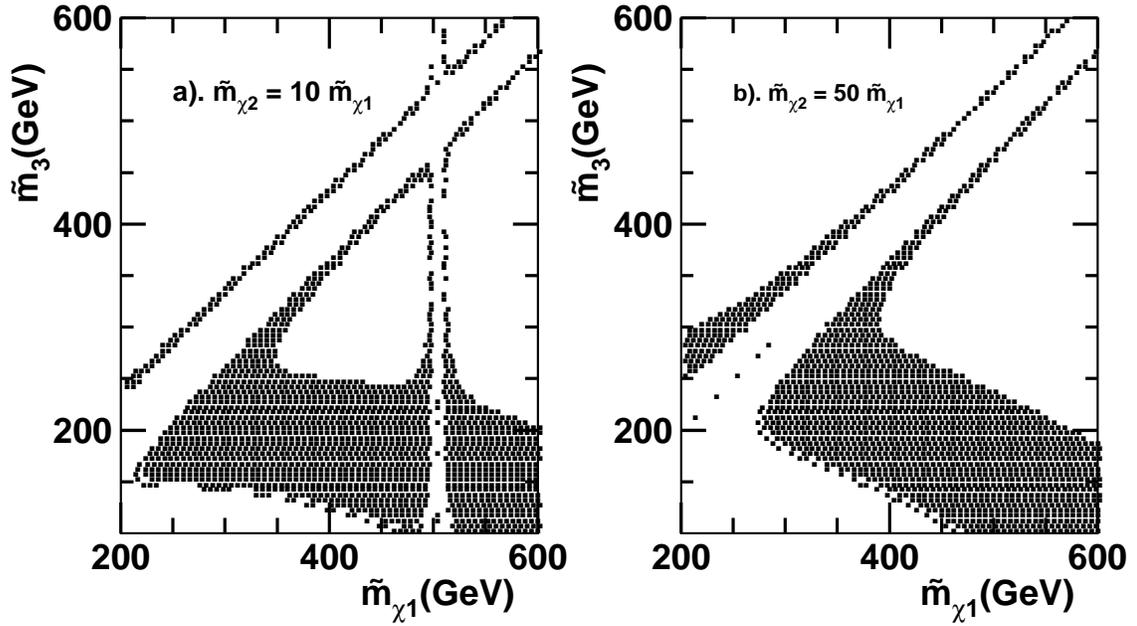}
\caption{In LL + RR  mixing case. $S_{\phi K_S}$ in the plane of $\tilde{m}_{L3}$ vs $\tilde{m}_{\chi_1}$. The shaded region corresponds to $ -0.7 < S_{\phi K_S} < 0.7$ with $\phi = 0.01$, $\theta = \pi/4$ and $\kappa = -1.1$. (a) $\tilde{m}_{\chi 2} = 10 \tilde{m}_{\chi 1}$; (b) $\tilde{m}_{\chi 2} = 50 \tilde{m}_{\chi 1}$;  
}
\label{lr:chi}
\end{figure}
\newpage
\begin{figure}
\dofig{15cm}{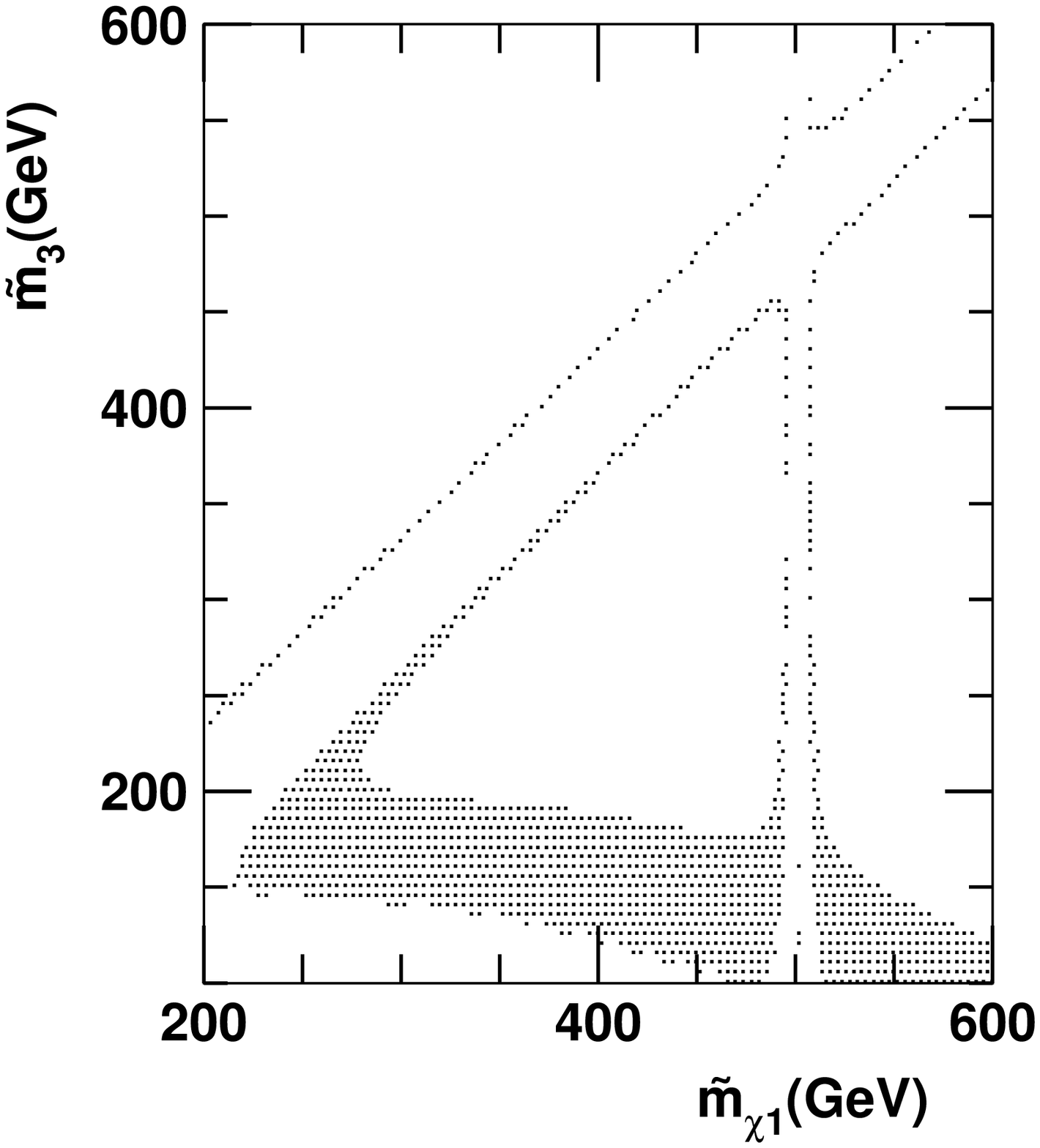}
\caption{LL + RR case.  $S_{\phi K_S}$ in the plane of $\tilde{m}_{L3}$ vs $\tilde{m}_{\chi_1}$. The shaded region corresponds to $ -0.7 < S_{\phi K_S} < 0.7$ with $\phi = 0.005$, $\theta = \pi/4$ and $\kappa = -1.1$. $\tilde{m}_{\chi 2} = 10 \tilde{m}_{\chi 1}$
}
\label{lr:phi}
\end{figure}
\newpage

\begin{figure}
\dofig{15cm}{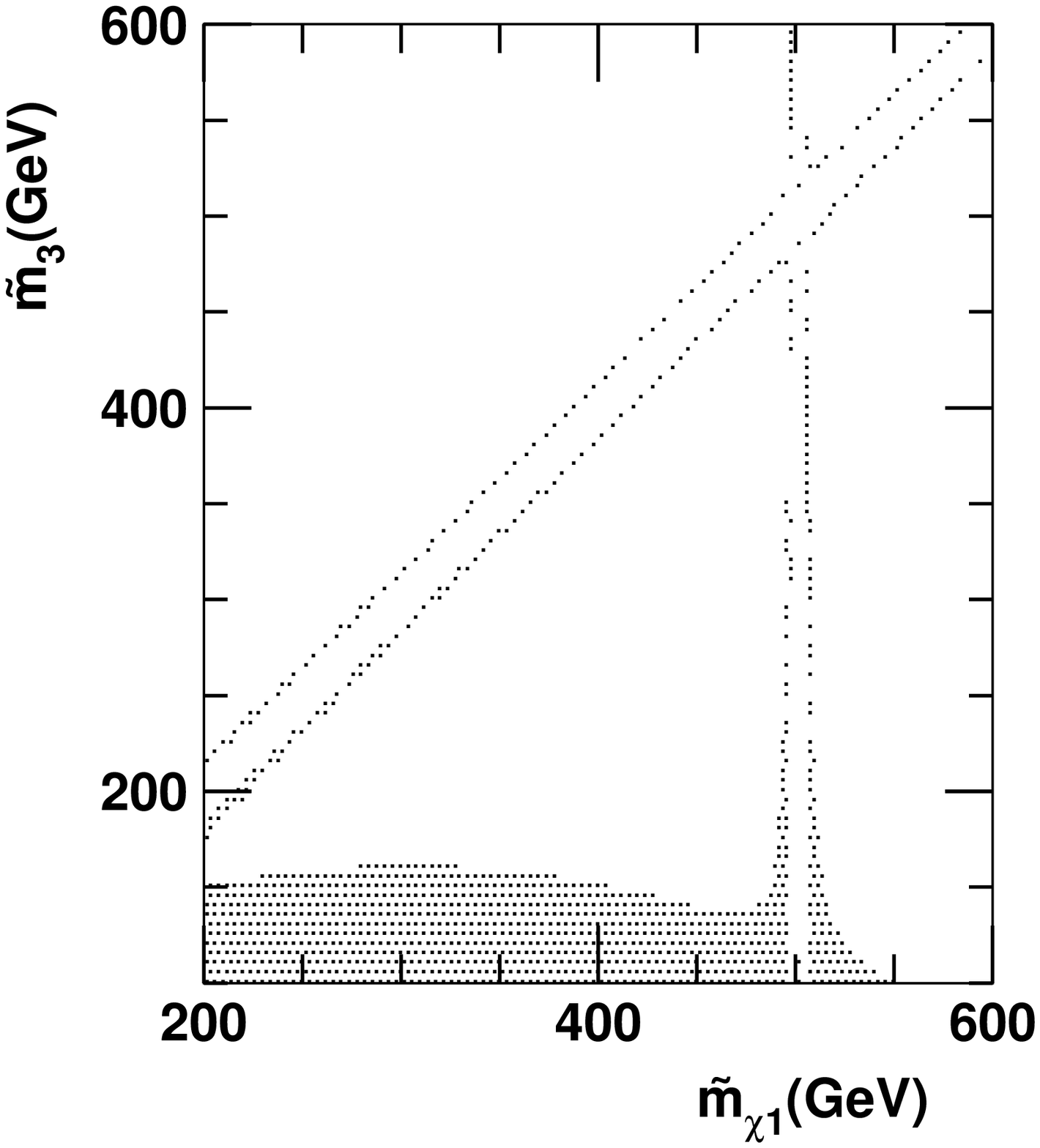}
\caption{LL + RR case. $S_{\phi K_S}$ in the plane of $\tilde{m}_{L3}$ vs $\tilde{m}_{\chi_1}$. The shaded region corresponds to $ -0.7 < S_{\phi K_S} < 0.7$ with $\phi = 0.01$, $\theta = \pi/6$ and $\kappa = -1.1$. $\tilde{m}_{\chi 2} = 10 \tilde{m}_{\chi 1}$
}
\label{lr:theta}
\end{figure}
\newpage
\begin{figure}
\dofig{16cm}{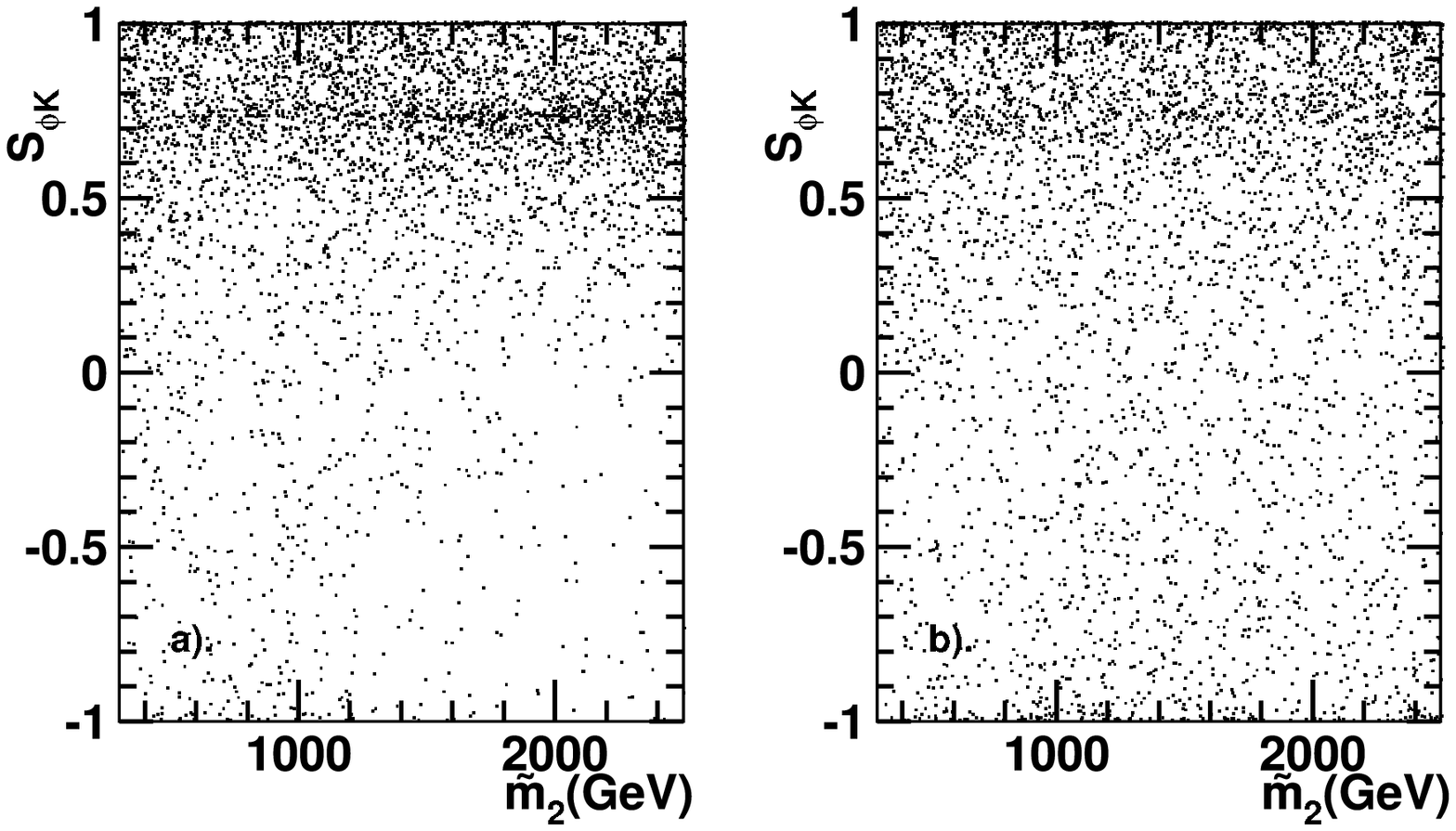}
\caption{LL + RR case. The CP asymmetry $S_{\phi K_S}$ scanned within parameter space over the region mentioned in the text. We show the results in the plane of  $S_{\phi K_S}$ vs $\tilde{m}_{L3}$ with (a). tan$\beta$ = 15 and (b). tan$\beta$ = 50}
\label{lr:scan}
\end{figure}
\newpage


\end{document}